\documentclass{article}

\usepackage[preprint]{neurips_2024}

\usepackage{tikz}
\usetikzlibrary{shapes,arrows,positioning}

\usepackage[utf8]{inputenc} 
\usepackage[T1]{fontenc}    
\usepackage{hyperref}       
\usepackage{url}            
\usepackage{booktabs}       
\usepackage{amsfonts}       
\usepackage{nicefrac}       
\usepackage{microtype}      
\usepackage{xcolor}         

\title{Accelerating Malware Classification: A Vision Transformer Solution}

\author{%
  Shrey Bavishi\\
  Department of Computer Science\\
  Indian Institute of Technology-Bombay\\
  Mumbai, India 400076 \\
  \texttt{200050132@iitb.ac.in} \\
  \And
  Shrey Modi \\
  Department of Computer Science\\
  Indian Institute of Technology-Bombay\\
  Mumbai, India 400076 \\
  \texttt{200020135@iitb.ac.in} \\
}

\begin{document}

\maketitle

\begin{abstract}
  The escalating frequency and scale of recent malware attacks underscore the urgent need for swift and precise malware classification in the ever-evolving cybersecurity landscape. Key challenges include accurately categorizing closely related malware families. To tackle this evolving threat landscape, this paper proposes a novel architecture LeViT-MC which produces state-of-the-art results in malware detection and classification. LeViT-MC leverages a vision transformer-based architecture, an image-based visualization approach, and advanced transfer learning techniques. Experimental results on multi-class malware classification using the MaleVis dataset indicate LeViT-MC's significant advantage over existing models. This study underscores the critical importance of combining image-based and transfer learning techniques, with vision transformers at the forefront of the ongoing battle against evolving cyber threats. We propose a novel architecture LeViT-MC which not only achieves state of the art results on image classification but is also more time efficient.

\end{abstract}

\section{Introduction and Background}
\label{sec:introduction}
The modern cybersecurity landscape is fraught with an ever-increasing threat
posed by malware, presenting a relentless challenge to security professionals and
researchers. Despite tireless efforts within the cybersecurity industry to combat these threats, cyber attackers remain undeterred, continually evolving their
tactics and techniques, and devising sophisticated evasive strategies such as polymorphism, metamorphism, and code obfuscations which make the problem very challenging.



This paper introduces LeViT-MC, an innovative approach to malware characterization and analysis utilizing the MaleVis dataset. Our methodology builds upon the insights of \citet{paper}, which suggest that malware executables can be effectively represented as image-like matrices, revealing significant visual similarities among malware from the same family.

In this study, we will first review the various methods previously employed for malware detection and classification, along with their inherent limitations. Subsequently, in \hyperref[sec:method]{Section \ref{sec:method}} , we will propose our novel architecture, LeViT-MC, which leverages the binary classification capabilities of DenseNet and the rapid inference speed of vision transformers. To the best of our knowledge, this represents the first instance of combining DenseNet and vision transformer architectures for the purposes of malware detection and classification.
\hyperref[sec:exp]{Section \ref{sec:exp}} details the experimental procedures and results obtained, demonstrating state-of-the-art accuracy and inference times attributable to our novel architecture. We will discuss the limitations of our work in \hyperref[sec:limit]{Section \ref{sec:limit}} , before concluding in \hyperref[sec:conc]{Section \ref{sec:conc}}. All code utilized to generate the results discussed in this paper can be found at this \href{https://github.com/Shrey-55/MalwareClassification}{link}.

\subsection{Static and Dynamic Analysis}

Early malware detection methods primarily relied on static and signature-based approaches. \citet{schul} explored the extraction of static features, including byte sequences and \citet{10.1016/j.ins.2011.08.020} explored the opcode patterns, in conjunction with machine learning techniques for classification. Dynamic analysis methods aimed to understand malware behavior by monitoring network activities and system calls. \citet{7491575} introduced a malware classification approach based on Hidden Markov Models (HMMs), analyzing API call sequences. However, dynamic analysis proved inefficient, particularly when malware adapted its behavior during execution.


\subsection{Vision-Based Approaches}
The transition to vision-based approaches marked a significant advancement in malware detection. \citet{10.1007/978-981-15-4018-9_6, Singh2019MalwareCU, 8888406} began visualizing features like opcode sequences and system calls as images, offering new perspectives on classification. \citet{paper} developed an efficient approach to visualising binary files as greyscale images and classfied the images using K-nearest neighbours. \citet{Conti2008VisualRE} demonstrated the effectiveness of visual methods in classifying binary files and analyzing new file structures. 

In their follow-up study, \citet{FALANA20221968} introduced a malware detection system achieving an impressive 99.8\% accuracy on malware detection. These methods talk only about malware detection, that is, a binary classification into beningn and malign classes, but fail to work in detecting the type of malware, which is also very crucial for system security.

\subsection{Transformer-Based Model}
Recent years have witnessed the rise of transformer-based models in malware detection. \cite{10.1145/3338501.3357374}, \cite{rahali2021malbert} introduced BERT(Bidirectional Encoder Representations from Transformers)-based models, 
\cite{mclaughlin2022malceiver} presented Malceiver, a hierarchical Perceiver model and \citet{9343222} introduced hierarchical transformer architectures for malware classification. The application of transformer-based models was expanded beyond assembly code analysis by Sherlock(\cite{Seneviratne_2022}) a transformer-based model for Android malware classification. Most of these methods rely on analysing and integrating opcodes, which significantly increases the inference time for detection and classification.





\section{Methodology}
\label{sec:method}
\subsection{Image Representation of Malware}
Our approach involves transforming PE binary files into RGB images by reading groups of 3 bytes and arranging them in a 2-dimensional vector space. Each byte, ranging from 0 to 255, is visualized as a pixel value in an image. The three bytes constitute one pixel in each of the three channels (RGB). This method preserves location information between malicious attack patterns and captures the order of these patterns, which is essential for precise classification, especially when dealing with structurally similar malware instances.
\subsection{ LeViT-MC Architecture}
\label{sec:arch}
We propose a novel architecture, LeViT-MC, that combines the strengths of Convolutional Neural Networks (CNNs) and Vision Transformers (ViT) to create a robust malware classification system. The architecture consists of two main components:
Binary Classification Stage: A fine-tuned DenseNet CNN followed by a classification head categorizes input images into benign and malign.
Malware Family Classification Stage: For images classified as malign, a LeViT (Lightweight Vision Transformer) further classifies them into specific malware families.
\subsubsection*{Binary Classification Stage}
We utilize a fine-tuned DenseNet CNN for the initial binary classification of images into benign and malign categories. DenseNet has demonstrated excellent performance in binary classification tasks for malware images as also showed by \citet{FALANA20221968} . The dense connectivity pattern in DenseNet allows for better feature reuse and improved information flow, making it particularly effective for capturing the intricate patterns present in malware images.
\subsubsection*{ Malware Family Classification Stage}
For the more granular task of classifying malign images into specific malware families, we employ the LeViT architecture  introduced by \cite{graham2021levit}. LeViT is designed for faster inference in image classification tasks, making it particularly suitable for real-time malware detection scenarios. More details about the LeViT architecture are given in \hyperref[sec:appenA]{Appendix A}

\subsection{Transfer Learning and Fine-tuning}
To leverage the power of pre-trained models and adapt them to our specific task, we employ transfer learning techniques:
We initialize the DenseNet component with weights pre-trained on ImageNet.
The LeViT component is initialized with weights from a model pre-trained on a large-scale image classification task.
We fine-tune both components on the MaleVis(\citet{malevis}) dataset, allowing the model to adapt to the specific characteristics of malware images.
This approach enables our model to benefit from the general feature extraction capabilities learned from large datasets while specializing in the nuances of malware classification.

\begin{figure}[h]
\label{img:workflow}
\centering
\begin{tikzpicture}[node distance=0.9cm, auto,
    block/.style={rectangle, draw, fill=blue!20, 
    text width=5em, text centered, rounded corners, minimum height=4em},
    line/.style={draw, -latex'}]

    \node [block] (file) {Input File};
    \node [block, right=of file] (input) {Transformed Image};
    \node [block, right=of input] (densenet) {DenseNet CNN};
    \node [block, right=of densenet] (binary) {Binary Classification};
    \node [block, below=of binary] (levit) {LeViT};
    \node [block, right=of levit] (family) {Malware Family Classification};
    
    \path [line] (file) -- (input);
    \path [line] (input) -- (densenet);
    \path [line] (densenet) -- (binary);
    \path [line] (binary) -- node [near start] {Malign} (levit);
    \path [line] (levit) -- (family);
    \path [line] (binary) -- node [above] {Benign} ++(1,1);

\end{tikzpicture}
\\

\caption{LeViT-MC Workflow}
\end{figure}
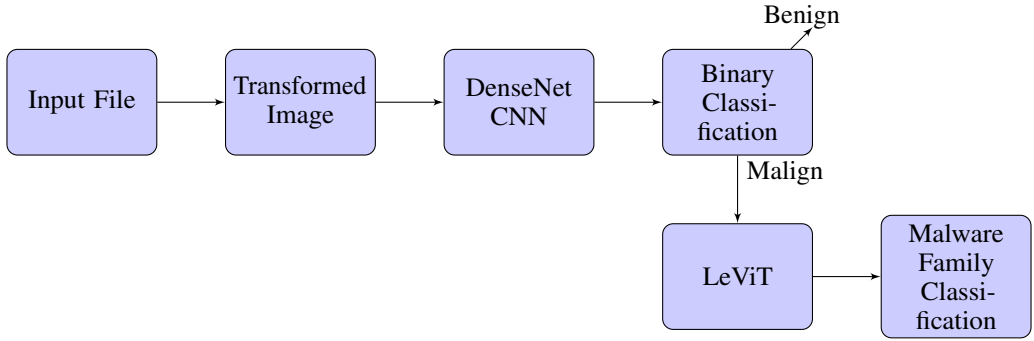

A complete workflow of our architecture is given in \hyperref[img:workflow]{Figure 1}
\section{Experiments}
\label{sec:exp}
\subsection{Dataset and Implementation Details}
We utilized the MaleVis(\citet{malevis}) dataset, comprising 14,226 RGB images in 224x224 pixel resolution. The dataset is divided into 26 categories: one representing benign samples and 25 representing various malware types. We have included some samples and more information about the dataset in \hyperref[sec:appenB]{Appendix B}. We employed the default partition of 70\% training and 30\% validation data.
Experiments were conducted using an NVIDIA A100-SXM4 GPU with 80GB RAM. Training was performed with a batch size of 32, using the Adam optimizer with an initial learning rate of 1e-5 and a ReduceLROnPlateau scheduler with a decay of 0.1 and patience of 10.
\subsection{Results}
LeViT-MC demonstrated superior performance in both accuracy and inference speed. We achieved 96.6\% accuracy, a groundbreaking multiclass classification accuracy on the MaleVis dataset, outperforming all the  previous state-of-the-art models. A comparision with previous multiclass state-of-the-art classification accuraries is given in \hyperref[tab:accuracy]{Table 1}

\begin{table}
  \caption{Accuracy comparison with various state-of-the-art models}
  \label{sample-table}
  \centering
  \begin{tabular}{lll}
   
    \cmidrule(r){1-2}
    Study       & Accuracy \\
     \hline
        \citet{paper} & 91.69\% \\
        \citet{agarap2019building} & 79.36\% \\
        \citet{a14100297} & 93\% \\
        \citet{10.1093/comjnl/bxac181}  & 93.49\% \\
        \textbf{LeViT-MC} & \textbf{96.6\%} \\
        \hline
    \bottomrule
  \end{tabular}
\end{table}

While gaining the highest accuracy, our novel architecture also gains the highest inference speed by utilising the fast inference of LeViT transformers. An inference speed comparision with the average inference speed for different architectures on the same dataset as quoted in the study by  \citet{Bianco_2018} is shown in \hyperref[fig:perfcomp]{Figure 2}. Our model is about has around 10x better inference times than normal vision transformers and around 3x better inference times than the best CNNs like ResNet.
\begin{figure}[h]
    \centering
    \includegraphics[width = 0.7\linewidth]{ 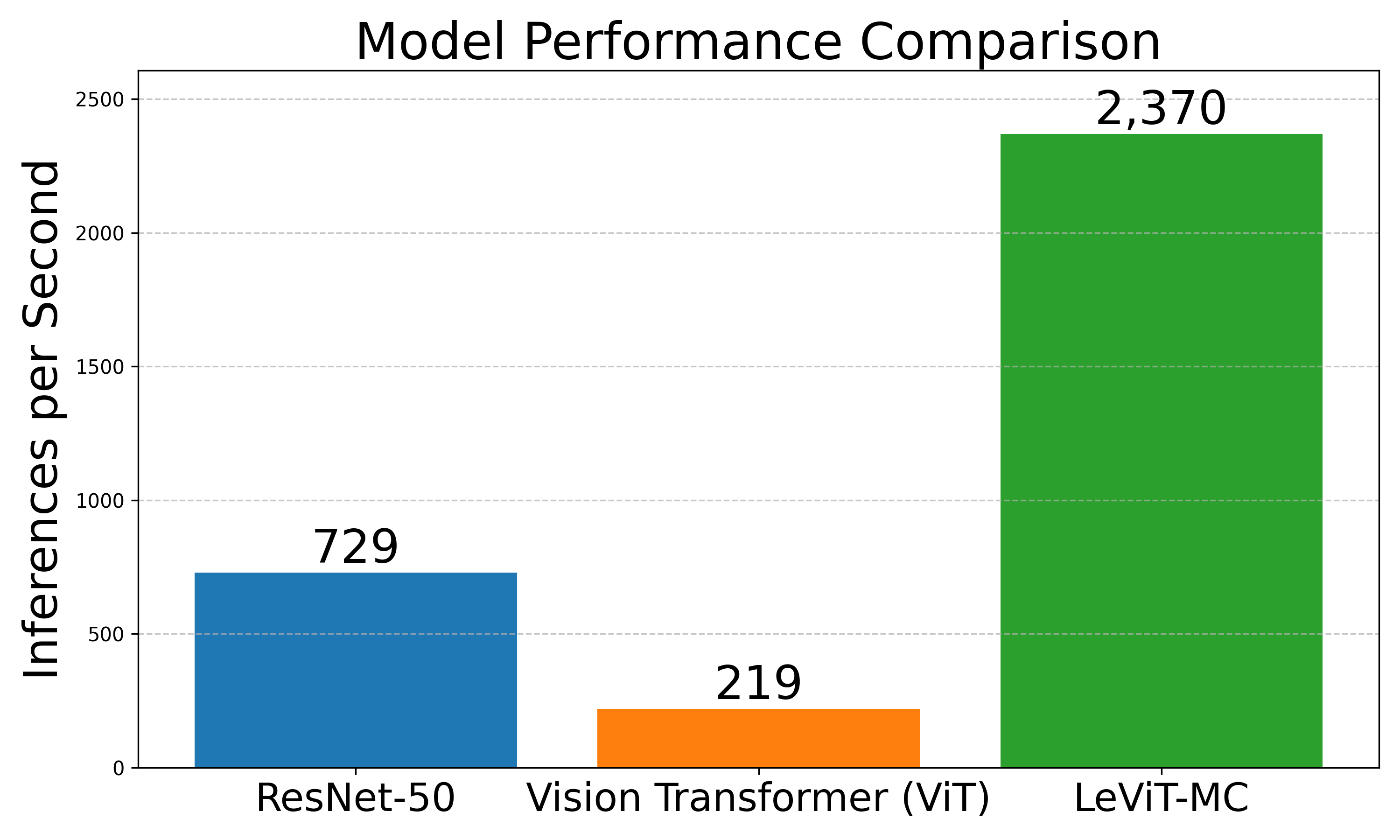}
    
    \caption{Perfomance Comparision}
    \label{fig:perfcomp}
\end{figure}
\section{Limitations}
\label{sec:limit}
While LeViT-MC has demonstrated exceptional performance, a couple of considerations remain:
\begin{itemize}
    \item \textbf{Dataset Constraints:} Our current evaluation was performed on the MaleVis dataset, and while this dataset covers a broad spectrum of malware types, its generalizability to other datasets remains untested.
    \item \textbf{Real-World Deployment and Scalability:} The computational efficiency observed in controlled environments (e.g., with high-performance GPUs) may not fully translate to resource-constrained devices like IoT systems or embedded malware detection platforms.
\end{itemize}

\section{Conclusion}
\label{sec:conc}
LeViT-MC demonstrates exceptional performance in malware classification, achieving 96.6\% accuracy and processing 2370 images per second. These results underscore the potential of combining CNNs and lightweight Vision Transformers in addressing the challenges of rapid and accurate malware detection and classification.

Looking ahead, our future work will emphasize the practical deployment of our methodology in real-world scenarios. Effective implementation in cybersecurity infrastructures is essential to proactively defend against evolving malware threats. We aim to explore integration with existing security systems, ensuring that our approach not only meets performance benchmarks but also adapts to the dynamic nature of cyber threats. Additionally, we will focus on enhancing explainability in our model to foster trust and transparency in its decision-making processes, facilitating its acceptance in operational environments.

\newpage

\bibliographystyle{plainnat}
\bibliography{ref}


\newpage
\appendix

\section{Appendix A: The LeViT Architecture}
\label{sec:appenA}

The LeViT-256 architecture, part of the LeViT family, represents a hybrid approach that combines the strengths of convolutional networks and transformer models for image classification tasks. This architecture addresses the computational inefficiencies of standard transformers by integrating convolutional stages to enhance spatial inductive biases, resulting in improved performance and reduced complexity.

LeViT-256 employs a multi-stage structure consisting of:

\begin{enumerate}
    \item A convolutional stem for efficient initial image processing and dimension reduction
    \item Transformer blocks for capturing global dependencies
    \item Attention pooling for effective information aggregation
\end{enumerate}

This design enables LeViT-256 to achieve a balance between accuracy and computational efficiency, making it suitable for low-latency and edge-based deployment scenarios. The architecture processes images through hierarchical down-sampling within its transformer layers, reducing the overall number of operations while maintaining strong performance on image classification benchmarks.

By leveraging multi-scale feature extraction and maintaining smaller feature map sizes, LeViT-256 significantly reduces computational overhead. The model transforms images into patches, processes them through stages of attention layers that capture global contextual features while reducing image resolution, and finally classifies the output using a linear head. This approach allows LeViT-256 to achieve state-of-the-art accuracy while maintaining high computational efficiency, positioning it as a promising solution for real-time image classification applications.
\\
\begin{figure}[h]
    \centering
    \includegraphics[width = 0.8\linewidth]{ 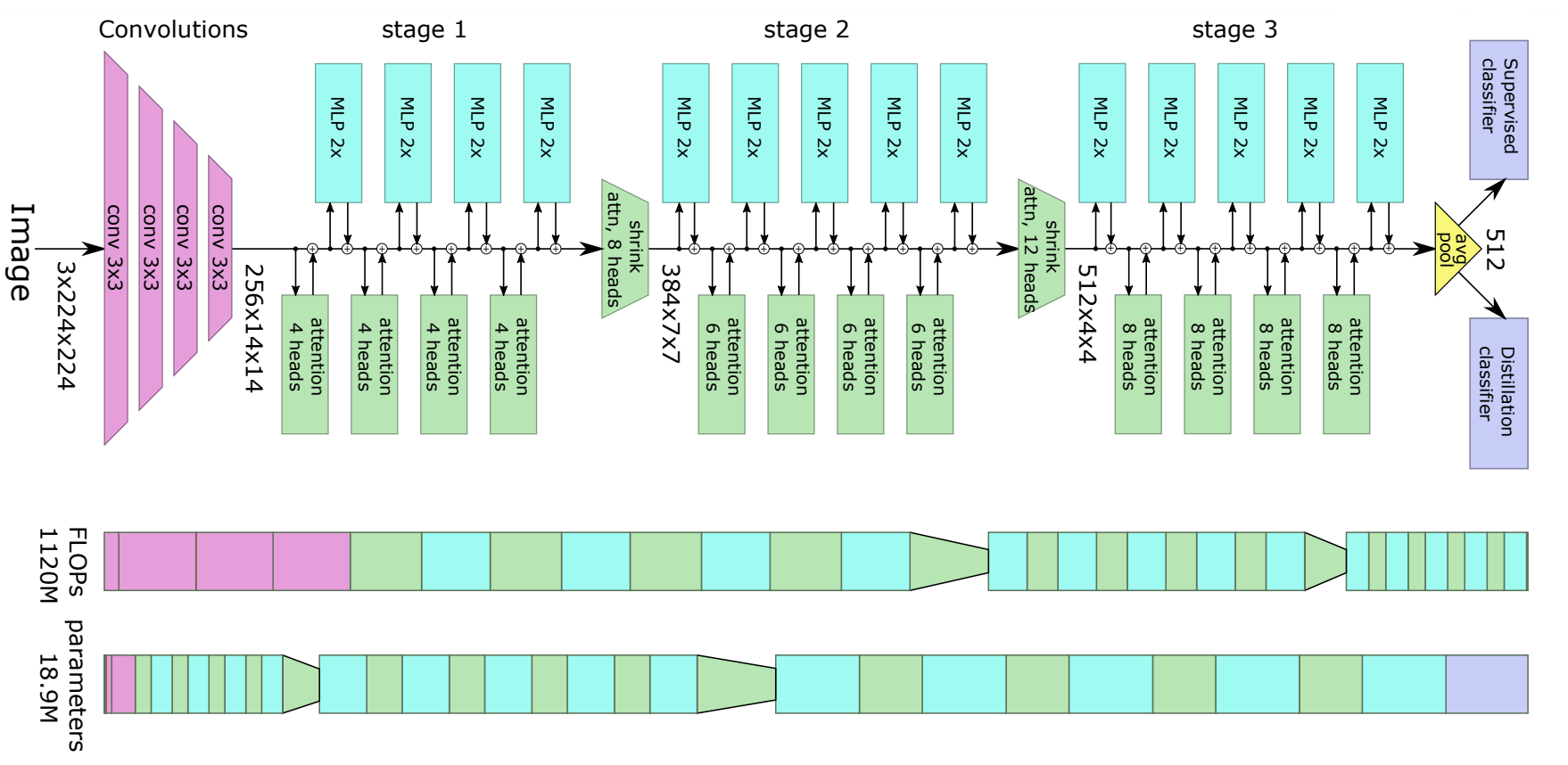}
    \caption{LeViT-256 Architecture}
    \label{fig:perfcomp}
\end{figure}

\section{Appendix B: MaleVis Dataset}
\label{sec:appenB}
In our experimental setup, we employed the MaleVis dataset, short for ”Malware Evaluation with Vision,” which was publicly released in 2019. This dataset comprises a total of 14,226 images, provided in two square resolutions, specifically 224 by 224 pixels and 300 by 300 pixel, that have been converted to the RGB format. The dataset is divided into 26 distinct categories, with one representing benign samples and the remaining 25 representing various types of malware. Our study utilized images with a 224 by 224 pixels resolution. As illustrated in Figure
5, the dataset exhibits balanced class distribution among various malware types each comprising approximately 500 images. In contrast, the ”Normal” class contains more samples, totalling 1832 images. To conduct our experiments, we use the default partition given by the dataset, containing 70\% images in the train dataset and 30\% in the validation dataset.

\begin{figure}[h]
    \centering
    \includegraphics[width = 0.5\linewidth]{ 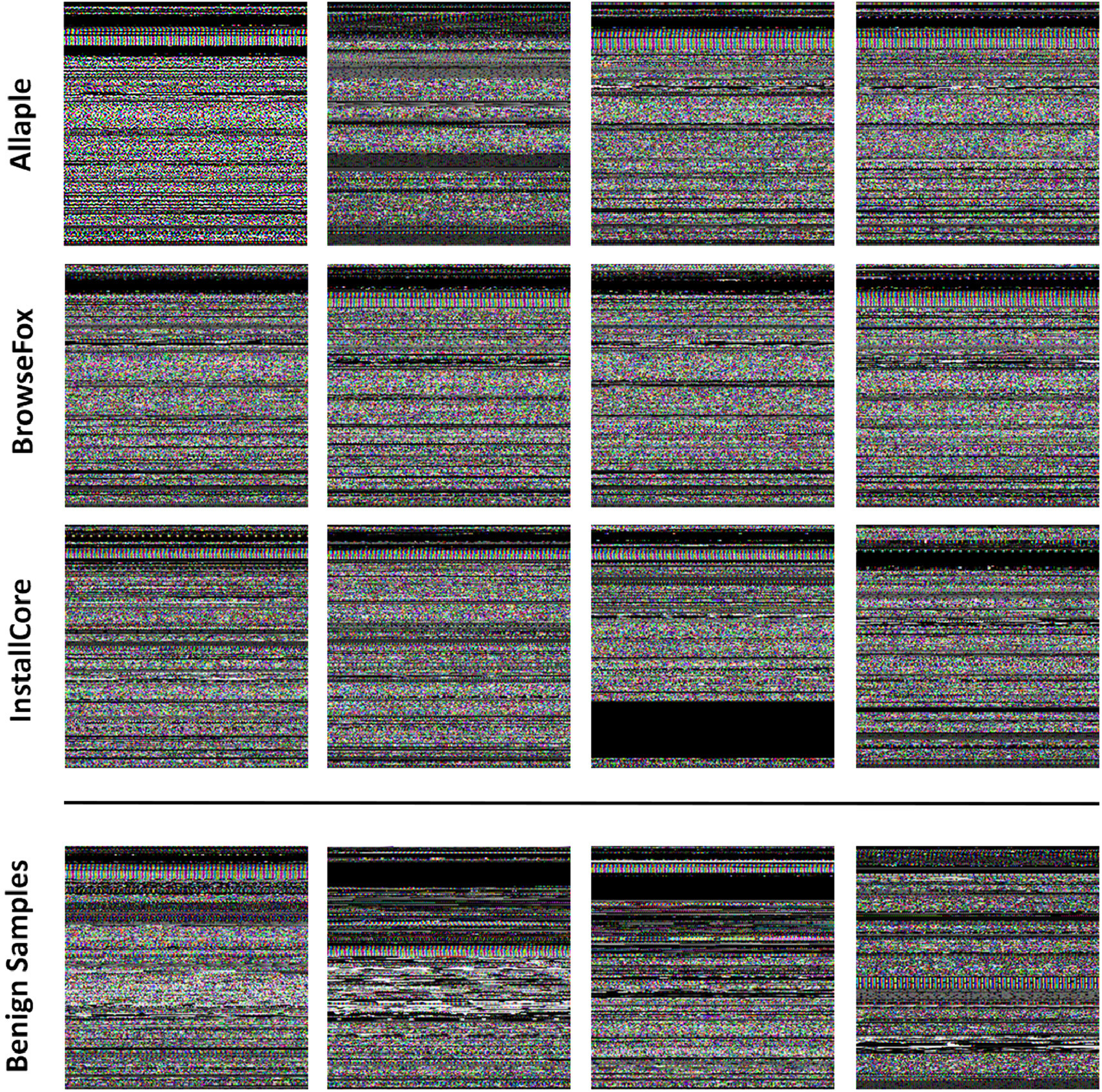}
    \caption{Sample images from the MaleVis Dataset}
    \label{fig:perfcomp}
\end{figure}

\begin{figure}[h]
    \centering
    \includegraphics[width = 0.8\linewidth]{ 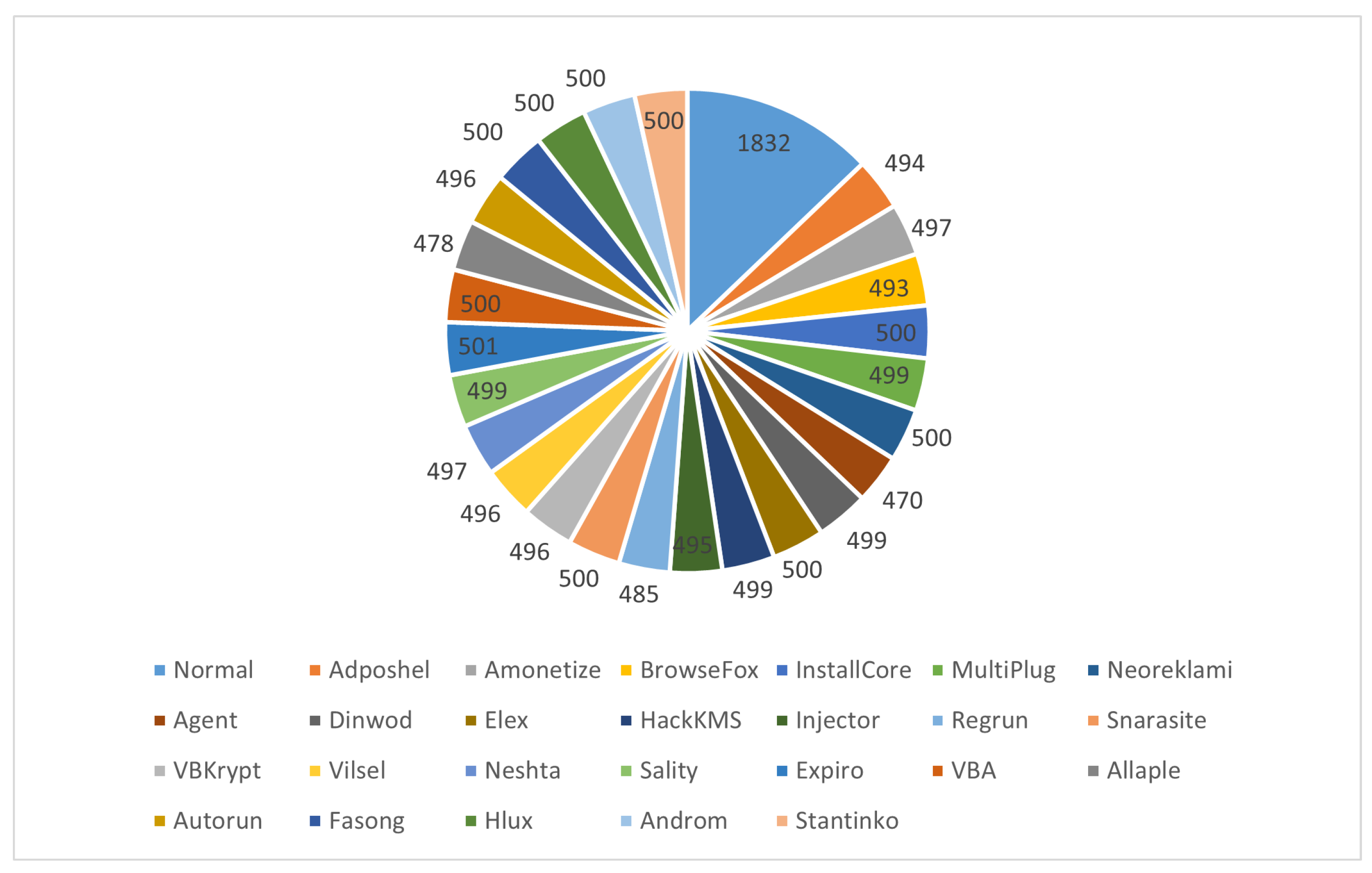}
    \caption{Distribution of the various classes of malwares}
    \label{fig:perfcomp}
\end{figure}


\newpage

\end{document}